# Overview of Spin-based Quantum Dot Quantum Computation

**Xuedong Hu**[*,1,2,3] **and S. Das Sarma**[1]

[1] Condensed Matter Theory Center, Department of Physics, University of Maryland, College park, MD 20742-4111, USA

[2] Current address: Digital Materials Lab, Frontier Research System, RIKEN, 2-1 Hirosawa, Wako-shi, Saitama, 351-0198, JAPAN

[3] Permanent address: Department of Physics, University at Buffalo, SUNY, Buffalo, NY 14260, USA

Here we give an overview of the research on spin-based quantum dot quantum computation. In particular, we discuss the conditions for the Heisenberg exchange Hamiltonian to be a satisfactory description of two-electron interaction in a GaAs double dot. We also discuss theoretical issues related to spin relaxation, spin encoding, and spin measurement. A brief account is also given to spin-based quantum computation in SiGe quantum dots and donors in Si.

**1 Introduction** During the past decade, the study of quantum computing and quantum information processing has generated great interest among physicists from atomic physics to optics to condensed matter physics. The intrinsic parallelism in quantum mechanics due to superposition and quantum correlations has been shown to lead to greater computational power for a quantum computer (QC). The most striking examples are the exponential speed up (by a QC compared to the best classical algorithms in a classical computer) in factorization using Shor's algorithm [1] and the square root speed up using Grover's algorithm [2]. When quantum error correction was shown to be possible [3], researchers started to look for physical systems that are appropriate as quantum bit (qubit). Some of the more prominent examples include trapped ions [4], cavity QED systems [5], liquid state NMR [6], electron spins in semiconductor quantum dots [7], donor nuclear spins [8,9], superconductors [10], etc. The experimental demonstration of these hardware proposals can be characterized as very fruitful in some cases, such as trapped ions, and very difficult in some other cases, such as most solid state based schemes.

With all the difficulties in demonstrating quantum coherence and controlled entanglement in solid state systems, especially spins in semiconductor systems, they are still often considered the most promising candidates for making large quantum computers because of the belief that they are scalable. Such a belief is largely based on the unqualified successes of the semiconductor industry during the past half a century and the immense resource available from this industry. Physically, spins can possess very long decoherence time even in solids, making them a good candidate for a qubit. Technologically, miniaturization of electronic devices has been a continuing success and has been rapidly improving. Here we review some of the spin-based QC schemes that have been proposed, and the theoretical and experimental research progresses.

**2 Spin-based quantum computers** A fermionic spin (more specifically, spin-1/2), being a quantum two-level system in a magnetic field, is a natural qubit with its spin-up and spin-down states. For example, the most successful demonstration of quantum control and entanglement is in a trapped ion system

[*] Corresponding author: xhu@buffalo.edu, Phone: +1 716 645 2017, Fax: +1 716 645 2507

using two hyperfine split levels for qubit [11]. These levels are essentially two nuclear spin levels. It was also proposed in the mid-1990s that Ising interaction between electron spins in solids can produce the desired entanglement for a QC [12]. One of the most influential spin-based solid state quantum computer schemes is based on exchange interaction in a double quantum dot [7]. As is illustrated in Fig. 1, the qubit in this scheme is the spin of the single electron trapped in each quantum dot. Linear arrays of such quantum dots form the backbone of this QC architecture. Single-qubit operations can be achieved by local magnetic fields or g-factor engineering, which create locally different Zeeman splitting for the electron spin. Two-qubit operations are based on the exchange interaction in a double quantum dot, where the inter-dot potential barrier can be lowered to increase electron wavefunction overlap, thus tuning the exchange interaction between the electrons. The form of this interaction is argued to be Heisenberg spin exchange: $J\,\mathbf{S}_1\cdot\mathbf{S}_2$, which means the orbital degrees of freedom of the electrons are all frozen. Spin measurement can presumably be accomplished by so-called spin valves made from ferromagnetic materials or spin blockade phenomena in electron transport [13]. In the following, we will first focus on the issues related to this quantum dot quantum computer architecture, and then briefly summarize the development of other spin-based QC schemes.

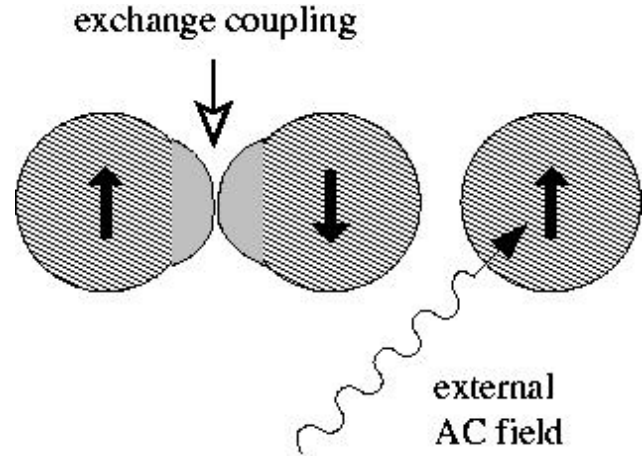

Figure 1. Schematics of an electron-spin-based quantum dot quantum computer architecture. The left two dots are interacting through increased overlap while the right dot is being manipulated by a local external field.

**3 Validity of Heisenberg exchange Hamiltonian in double quantum dot**  One of the key issues in the spin-based quantum dot quantum computer involving exchange coupling in a double dot is to show the applicability of Heisenberg spin exchange Hamiltonian to describe the two-electron interaction. Presumably, for small enough quantum dots, the orbital degrees of freedom is frozen, so that the two electron states in a double dot should be dominated by the ground spin singlet and triplet states (whose splitting is the so-called exchange splitting $J$). Another important issue is the order of magnitude of this exchange splitting $J$ and whether it is sufficiently large to support a practical device.

The theoretical calculation to characterize electron spin exchange in a double dot was first done using Heitler-London approach, in which the electron orbital states are limited to the ground states in the two single quantum dots that form the double dot [14]. The authors also included the doubly-occupied state (where both electrons are in the same dot) in the so-called Hund-Mulliken calculation to further improve the calculation of the exchange constant. They showed that the exchange constant can be sizable in a GaAs system.

We performed a molecular orbital calculation to further clarify the properties of the exchange splitting in a GaAs horizontal double quantum dot [15]. The system we considered is formed from two-dimensional electron gas by surface gate depletion [16]. The growth direction confinement is so strong that the energy scale along that direction is much higher than the horizontal direction and excitation along that direction is neglected (thus horizontal quantum dot). One reason to consider the horizontally coupled quantum dots is that the inter-dot coupling can be easily tuned with surface gate potential adjustments. In our calculation, we included the first excited orbital states (P orbitals) in the calculation to provide possibility of spatial deformation in the electron wavefunction. After all, molecular bonding is strongly affected by electron distribution in space, while the system we considered is essentially an effective two-dimensional hydrogen molecule. Our results showed that the inclusion of the P orbitals indeed affect the exchange coupling, generally causing an increase about 20% compared to the Hund-Mulliken calculation [15]. The size of the quantum dots we considered is quite small (in the order of 50 nm), somewhat smaller than the state of the art experimental value of about 100 nm. The increase in size invariably leads to smaller on-site Coulomb repulsion, smaller single particle excitation energy, and generally smaller exchange coupling.

The spectroscopic results for double quantum dot [14,15] show that exchange coupling should be sufficiently strong to implement quantum computing operations. For example, an exchange splitting of 0.1 meV corresponds to a swap gate with duration as short as a few tens of picosecond. Compared to the electron spin decoherence time in the order of μs, such gating time is short enough for quantum error correction codes to work properly. By discussing only the lowest spin singlet and triplet states, we are implicitly assumed that the higher energy two electron states are all negligible. Such assumption is based on the electron states being manipulated adiabatically. On the other hand, the gate operations are limited from above by the electron spin decoherence time. It is thus imperative to determine whether the adiabatic requirement is a stringent condition or not. Such calculation was carried out within the Hund-Mulliken model by assuming a particular temporal shape of the exchange splitting in a double dot [17]. We also performed such calculation with our calculated two-electron energy spectra and wavefunctions [18]. Instead of assuming a particular temporal profile for the exchange coupling $J$, we assumed a temporal variation in the inter-dot barrier height, which is a quantity that can be easily tuned by external voltages applied to the surface gates. Our results show that errors due to non-adiabaticity decreases rapidly as the gate operations become longer, and the requirement for adiabatic operation is not overbearing on the spin-based quantum dot QC [18].

The calculations on quantum dot spectra and adiabatic manipulation discussed above justify the use of Heisenberg spin exchange Hamiltonian to describe the low energy dynamics in a double quantum dot and the related two-qubit operations in a spin-based quantum dot QC. In these calculations spin-orbit coupling has been neglected because of their small magnitude for the conduction electrons in bulk GaAs (near the bottom of the conduction band the electron wavefunction is mostly formed from the atomic S orbitals). However, the horizontal quantum dots are made from two-dimensional electron gas confined in heterostructures or quantum wells, where the sharp surfaces and asymmetry lead to strong Rashba type spin-orbit coupling, which in turn leads to finite anisotropic exchange in the spin interaction [19] in the form of **h**·($S_1$×$S_2$) plus higher order corrections. Here **h** is a vector determined by spin-orbit interaction. The inclusion of these terms does not take away the capability of quantum dots to perform quantum logic operations (indeed, there has been a proposal to use the anisotropic exchange coupling to perform quantum logic operations [20]). Nevertheless, they do add more complexity to the aesthetically simple and beautiful isotropic Heisenberg exchange coupling. Fortunately, two different groups of researchers have proved that by carefully choosing the temporal profile of the inter-dot coupling it is possible to largely eliminate the effects of the anisotropic exchange due to the spin-orbit interaction [21].

Even when the two-electron interaction in a double dot can be characterized by the Heisenberg spin exchange Hamiltonian, inhomogeneity in the single spin environment can still cause problems in the two-qubit quantum logic operations. For example, we proved that inhomogeneous Zeeman coupling leads to incomplete swap operations [22]. This means that swap cannot be accomplished by a single pulse of exchange gate anymore. Instead, several pulses (at least 3) have to be used for large inhomogeneity [23], while smaller inhomogeneities such as those due to trapped charges nearby can be and have to be corrected [22]. Interestingly, inhomogeneous Zeeman coupling can also be utilized for the purpose of qubit encoding [24], which leads to all-exchange logical operations (eliminating the need for local magnetic field and/or g-factor engineering) [25] that originate from the concept of decoherence free subspace [26].

In an effort to relax the requirement of spin-based quantum dot quantum computation, we also studied whether a multi-electron quantum dot can be used as a qubit [27]. In particular, we performed a configuration interaction calculation for a double dot with six electrons (3 per dot) to explore whether the low energy dynamics is still entirely dominated by spin dynamics. Our results showed that orbital level degeneracy can lead to the participation of multiple levels in the low energy dynamics, therefore cause serious complexity and difficulty in spin exchange. To solve this problem, external means such magnetic field or quantum dot deformation has to be applied to lift the orbital degeneracy so that the electron cloud in each quantum dot can again be described by an effective spin-1/2 entity.

**4 Decoherence in quantum dots**   For an electron spin to work as a qubit, it is imperative to carefully evaluate its decoherence properties. It has been known for a long time that electron spin in materials like bulk GaAs and Si has very long relaxation time [28]. More recently, in the context of spin-based quantum dot quantum computing, spin relaxation channels and relaxation times are again being carefully studied. For example, spin-orbit coupling and its resulting spin decoherence was quite thoroughly analyzed [29]. Hyperfine coupling between electron spins and the surrounding nuclear spins have also been explored [14,30]. In particular, we have focused on the electron spin decoherence due to spectral diffusion originating from the fluctuations in the nuclear spin magnetic field caused by nuclear spin dipolar interaction [30]. It is comforting to know that the bottom-line spin decoherence time for electron spins in GaAs can be as long as 100 μs. Spin decoherence and its measurement will continue to be carefully studied in various material systems as it is one of the most fundamental issues in spin-based quantum computation.

**5 Spin-based quantum computation in silicon**   Our discussion up to now has focused on the spin-based QC schemes with GaAs as the host material. Since all the isotopes of both Ga and As have finite nuclear spins, it is very difficult to identify certain nuclear spin species as qubits. On the other hand, if the host material is Si, whose $^{28}$Si isotope has vanishing nuclear spin, both electron spin and nuclear spin can be considered as candidate for qubit.

One of the best publicized silicon QC proposals is based on the Si:P system, with the nuclear spin of phosphorus donors ($^{31}$P) as qubit [8]. In a general sense, donors can be considered quantum dots as well, as the confinement for electrons are provided by the Coulomb potential of the extra protons at the donor nuclear sites. In the architecture proposed in Ref [8], as illustrated in Fig. 2(a), the qubits are the nuclear spins of the $^{31}$P donors. To perform selective single-qubit operations, one can vary the voltage on a surface gate (A gate) that tunes the hyperfine splitting of the corresponding donor under the gate, and then apply a resonant external AC field. For two-qubit operations, overlap in neighboring donor electron wavefunctions is introduced, which would then produce an effective exchange interaction between the nuclear spins that are interacting with the electrons through the contact hyperfine coupling. The measurement of spin information is done through the electrons. The nuclear spin information is first transferred to the electron spins, then the electron spins are measured. This proposal carries the advantages that the quantum information is stored in the extremely reliable donor nuclear spins, while the manipulation of these information is done through the donor electrons, which are more easily controlled but still possess long decoherence time. Furthermore, this proposal has the enticing aspect that it is closely related to the dominant silicon technology, which encourages the hope that scalability can be more easily achieved backed by the vast resources possessed by the semiconductor industry.

There have been active experimental and theoretical research efforts in attempting to fabricate the regular donor arrays in a Si:P system [31]. Two main experimental approaches are adopted, one using ion implantation by bombarding $^{31}$P atoms into crystalline silicon, the other using MBE to grow the system layer by layer. In the first approach, it has been suggested by single electron transistors can be used to monitor the presence of donors. Using the later approach, it has been shown that precise positioning of phosphorus donors into a linear array can be achieved on a Si [001] surface [32]. Apparently, much more experimental efforts have to be invested and more sophisticated technology in donor positioning and manipulation has to be invented for this QC architecture to be fabricated with precision. Theoretically, we have recently performed Heitler-London type of calculation for the donor electron exchange in bulk and strained silicon [33]. Silicon conduction band has six potential minima close to the X points of the silicon First Brillouin Zone, so that donor electron wavefunctions have to be expanded on the basis of the six Bloch functions at these points. Our results showed that the resulting inter-valley interference leads to strong atomic scale oscillations in the inter-donor electron exchange, which presents a significant difficulty in the control of two-qubit operations [33]. We also showed that uniaxial strain can partially alleviate this problem, but not completely removing it. Further theoretical and experimental works are underway to study how to overcome the problems presented here [34].

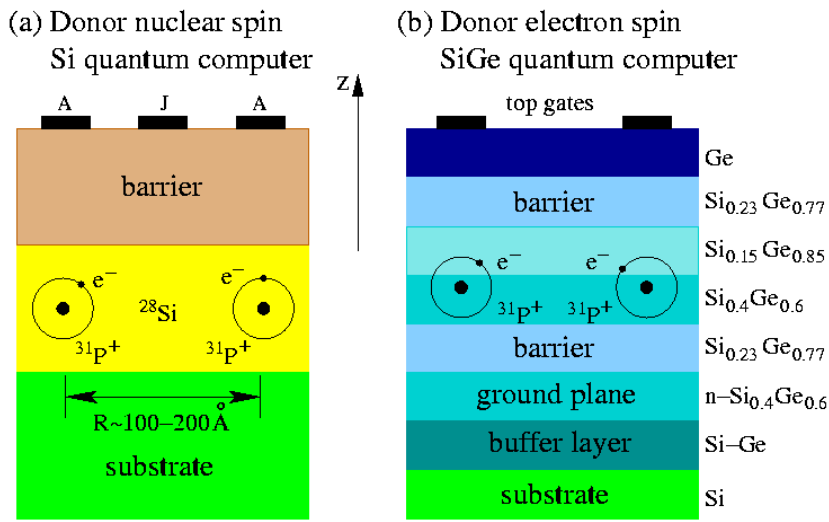

Figure 2. Schematics of donor nuclear and electron spin based quantum computer architectures.

There have also been proposals using electrons spins in Si or SiGe host materials for quantum computing. Figure 2(b) illustrates one such proposal, in which the spin of the donor electrons from $^{31}$P donors in SiGe alloy is the qubit [35]. By moving a donor electron into alloy regions of different g-factor, its Zeeman splitting can be tuned significantly, which then allows selective single-qubit operations. Similarly, different alloy regions also present different electron effective masses, which affect the electron wavefunction extension sensitively. Such property can then be used to tune the exchange coupling between two bound donor electrons. This all-electron proposal has speed-up gate operations compared to the nuclear-electron hybrid scheme mentioned above. However, the alloy environment has to be more thoroughly studied for such QC scheme to be realized.

A direct analogue to the electron-spin-based GaAs quantum dot quantum computer scheme has also been suggested [36]. Here the electrons are confined in the pure (but strained) silicon region (instead of SiGe alloys), while the confinement is produced by the SiGe alloys and surface gates. Compared to GaAs, the electron spin decoherence due to hyperfine coupling can be suppressed by using purified $^{28}$Si as the host material. On the hand, the more complicated band structure of silicon may pose problems to the coherent control of a quantum device, which needs to be further studied.

**6 Conclusions** In summary, there have been extensive research efforts in the spin-based quantum dot quantum computation. Theoretically, many important issues, such as control of spin interaction, spin decoherence, and spin measurement, have been carefully studied. Experimentally, a key road block, the single electron spin measurement, has yet to be overcome. A breakthrough in this technology would bring the field of spin-based solid state quantum information processing to a new height and much closer to its promise of scalable quantum computing.

**Acknowledgements** We acknowledgement financial support by ARDA, ONR, and LPS. XH also thank the hospitality of the Single Quantum Dynamics Research Group at the Frontier Research System, Riken, Japan.